\documentclass[doublecol]{epl2} 

\usepackage{graphicx,color,amsmath,amssymb}

\newcommand{\D}{\mathrm{d}}
\newcommand{\e}{\mathrm{e}}
\newcommand{\half}{\frac{1}{2}}
\newcommand{\be}{\begin{equation}}
\newcommand{\ee}{\end{equation}}
\newcommand{\bea}{\begin{eqnarray}}
\newcommand{\eea}{\end{eqnarray}}

\newcommand{\kbt}{k_{\mathrm{B}}T}

\newcommand{\lb}{\ell_\mathrm{B}}
\newcommand{\ld}{\lambda_\mathrm{D}}
\newcommand{\kd}{\kappa_\mathrm{D}}
\newcommand{\lgc}{\ell_\mathrm{GC}}

\begin{document}


\title{Charge regulation: a generalized boundary condition?}
\shorttitle{Charge regulation: a generalized boundary condition?}
\author{Tomer Markovich \inst{1} \and David Andelman \inst{1}\thanks{\email{andelman@post.tau.ac.il}} \and Rudi Podgornik \inst{2}}
\shortauthor{T. Markovich \etal}

\institute{ \inst{1} Raymond and Beverly Sackler School of Physics and Astronomy, Tel Aviv University, \\
Ramat Aviv, Tel Aviv 69978, Israel. \\
\inst{2} Department of Theoretical Physics, J. Stefan Institute, and Department of Physics, \\
Faculty of Mathematics and Physics, University of Ljubljana, 1000 Ljubljana, Slovenia.}
\date{October 19, 2015}

\abstract{
The three most commonly-used boundary conditions for charged colloidal systems are
constant charge (insulator), constant potential (conducting electrode) and charge regulation (ionizable groups at the surface).
It is usually believed that the charge regulation is a generalized boundary condition that reduces in some
specific limits to either constant charge or constant potential boundary conditions.
By computing the disjoining pressure between two symmetric planes for these three boundary
conditions, both numerically (for all inter-plate separations) and analytically (for small inter-plate separations),
we show that this is {\it not}, in general, the case.
In fact, the limit of charge regulation is a separate boundary condition,
yielding a disjoining pressure with a different characteristic separation-scaling.
Our findings are supported by several examples demonstrating that the disjoining pressure at small separations
for the charge regulation boundary-condition depends on the details of the dissociation/association process.
}

\maketitle

\section{Introduction}

Charge colloidal particles do not usually conform to the simple and popular idea that they can be characterized
either as insulators with fixed surface charges or conductors with constant surface potential~\cite{Adamson}.
In fact, when two colloidal particles with ionizable surface groups (immersed in an aqueous electrolyte solution) are brought together,
both their surface charge-density and surface electrostatic-potential change with the particle (surface) inter-distance~\cite{Borkovec1,Borkovec2}.
This ubiquitous phenomenon stems from the dissociation/association of surface ionizable groups and is referred to as {\it charge regulation} (CR).
It was elegantly formalized within the Poisson-Boltzmann (PB) theory of electrostatic interactions by Ninham and Parsegian in the 1970's~\cite{NP-regulation}.

The CR formalism can be implemented either through a chemical dissociation equilibrium of surface binding sites
(law of mass action)~\cite{Regulation2,Regulation3,Regulation4,Regulation5}, or through a surface-site partition function  (free energy)
\cite{Olvera,Olvera2,Natasa1,Natasa2,epl,diamant,maarten}. In both cases, it yields the same self-consistent boundary condition
for an effective surface-charge density that differs from the boundary condition of
constant charge (CC) for charged insulators or constant potential (CP) for conducting surfaces.
The concept of charge regulation has been widely applied in different situations:
analysis of the stability of the electrostatic double-layer and its relation to inter-surface forces \cite{stab,instab},
dissociation of amino acids and protein interactions~\cite{Leckband,Lund,Fernando},
charge regulation in protein aggregates such as viral shells \cite{Nap}
and polyelectrolytes and polyelectrolyte brushes~\cite{Netz-CR,Borukhov,Kilbey,Zhulina},
as well as for charged membranes~\cite{membranes1,membranes2,membranes3}.

Although the theory of charge-regulated electrostatic interactions has been previously used in numerous situations,
some conceptually important issues have not been addressed with sufficient generality.
Usually, the CR disjoining pressure, $\Pi_{_{\rm CR}}$, is bounded
by those stemming from the CC and CP boundary conditions~\cite{always}
(for some exceptions, see {\it e.g.}, Refs.~\cite{notalways1,notalways2}).
However, this {\it does not} imply that, in general, the expression of $\Pi_{_{\rm CR}}(d)$ as function of the inter-surface separation, $d$,
will properly reduce to the two implied limits.

In this Letter, we show on general grounds that the disjoining pressure, $\Pi_{_{\rm CR}}(d)$,
based on the CR boundary condition
has scaling properties in the limit of small inter-plate separations,
which differ from the scaling behavior of the CC or CP boundary conditions.

\section{Model}

Consider an ionic solution that contains monovalent
symmetric (1:1) salt of charge $\pm e$ of bulk concentration $n_b$,
immersed in aqueous solvent between two symmetric planes separated by distance $d$,
and of infinite lateral extent, as depicted in Fig~1.
We consider three types of boundary conditions: constant charge (CC), constant potential (CP) and charge regulation (CR).
The water solvent is assumed to be a continuum dielectric medium characterized by the water dielectric constant, $\varepsilon_w$.
We choose for convenience to locate the two planes at $z=\pm d/2$ such that $z=0$ is a symmetry plane.
Thus, the electrostatic potential is symmetric about the mid-plane, yielding a zero electric field, $E_m \propto \psi_m' = 0$ at $z = 0$.
\begin{figure}
\centering
\includegraphics[scale=0.4]{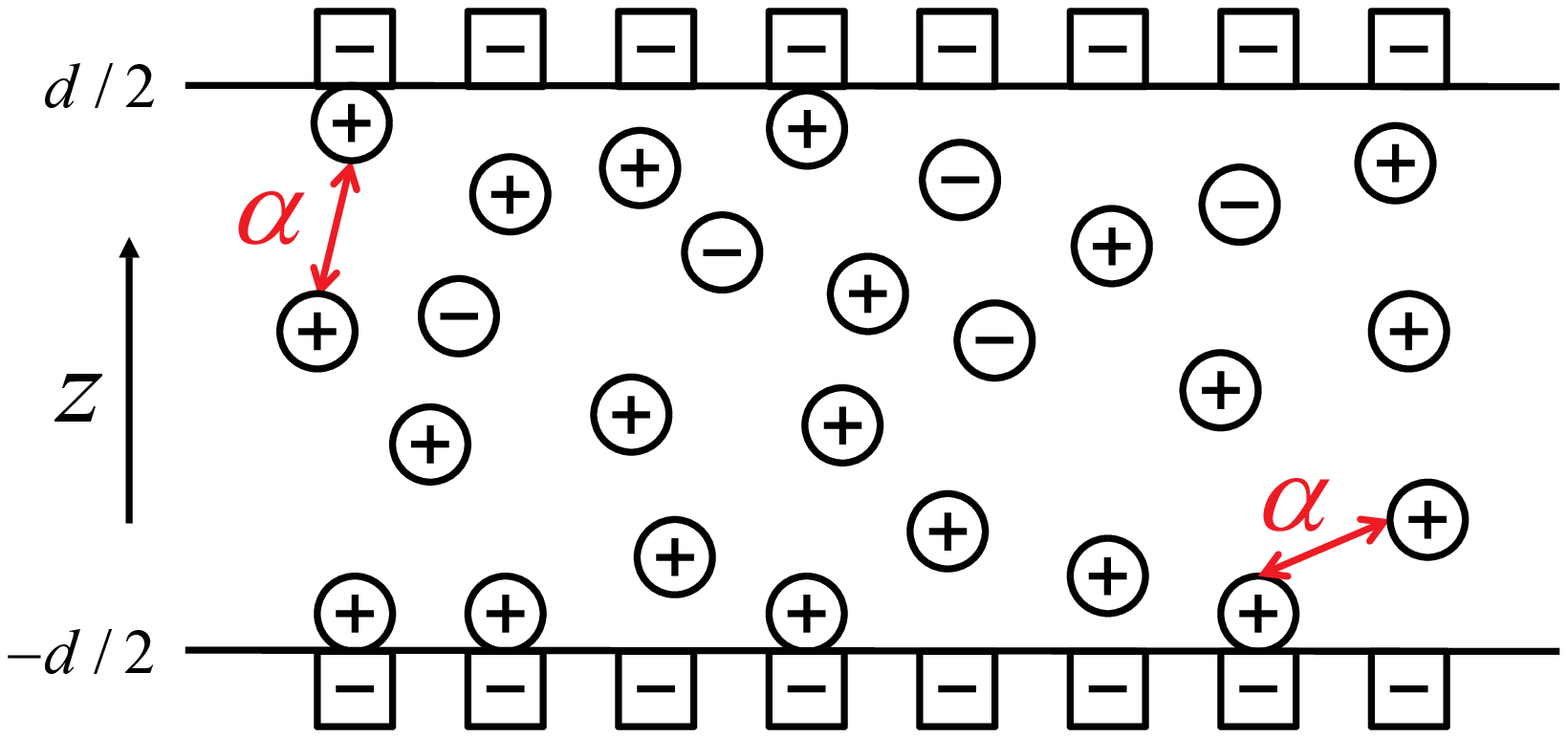}
\caption{ \textsf{Schematic drawing of two symmetric surfaces at $z=\pm d/2$ with dissociable ionic groups.
The charge regulation boundary-condition is described with a surface interaction parameter, $\alpha$.
The ions are dissolved in an aqueous solution of dielectric constant, $\varepsilon_w$.
  }
}
\label{figure1}
\end{figure}

The equation that governs the distribution of mobile ions in solution at finite temperature is the
well-known Poisson-Boltzmann (PB) equation (for details see Ref. \cite{Safinya}).
For 1:1 monovalent salts it has the simple form:
\begin{eqnarray}
\label{e1}
\frac{\D^2\psi}{\D z^2} = \kd^2 \sinh\psi(z) \, ,
\end{eqnarray}
where $\psi$ is redefined as a dimensionless electrostatic potential ($e \psi/\kbt \rightarrow \psi$)
and $\ld = \kd^{-1} = (8\pi e^2 n_b / \varepsilon_w \kbt)^{-1/2}$
is the Debye length, with $k_B$ the Boltzmann constant and $T$ the temperature.
This one-dimensional PB equation is obtained by taking into account the translation symmetry in the $x-y$ plane.

The solution  of the one-dimensional PB equation
can be expressed in terms of elliptic functions.
Exploiting the symmetry of the system, it is then sufficient to consider the interval $[0,d/2]$,
with $\psi_m' = 0$ at the $z=0$ mid-plane.
The general solution  in such a symmetric setup
can be written in terms of the Jacobi elliptic function~\cite{abramowitz,Safinya}, ${\rm cd}(u|a^2)$, as
\begin{eqnarray}
\label{e2}
\psi = \psi_m + 2\ln\left[ {\rm cd} \left( \frac{z}{2\ld\sqrt{m}} \Big| m^2 \right) \right] \, .
\end{eqnarray}
with $m \equiv \exp\left(\psi_m\right)$ and $\psi_m \equiv \psi(z=0)$.
The additional boundary condition at $z=d/2$ will determine a different $\psi_m$ for the three different cases: CC, CP, and CR at finite $d$.
Evaluating the above equation and considering the boundary condition at $z=d/2$
result in an explicit relation $\sigma = \sigma(\psi_s;d)$ between the surface charge density, $\sigma$,
the surface potential $\psi_s$ and $d$.
In order to understand how the three boundary conditions differ
and when they indeed merge, we use the general expression for the disjoining pressure,
which is valid in all three cases (CC, CP and CR) {as explained in Ref.~\cite{Harries}},
\begin{eqnarray}
\label{e2a}
\Pi(d) = 4\kbt n_b\sinh^2(\psi_m/2) > 0 \, .
\end{eqnarray}
This pressure is a macroscopic measurable quantity that strongly depends on the inter-plane separation, $d$.

The difference in the disjoining pressure for the three boundary conditions becomes substantial only for
relatively small separations, $d \lesssim \ld$,
while in the large separation limit, $d\ \gg \ld$, the three pressure expressions coincide.
To gain further insight into the different behavior of the disjoining pressure,
we focus on the small separation limit, $d \ll \ld$ and $d \ll \lgc$,
{where $\lgc \equiv e/(2\pi\lb|\sigma|)$ is the Gouy-Chapman length and $\lb = e^2/\varepsilon_w\kbt$ is the Bjerrum length}.
Our analytical results give the scaling of $\Pi$ with $d$ and clearly distinguish between the three boundary conditions.

Let us start with the most common CC boundary condition,
\begin{eqnarray}
\label{e3}
\psi^{\, \prime}\Big|_{z=d/2} \equiv \psi_s^{\, \prime} = 4\pi\lb \frac{\sigma}{e} \, .
\end{eqnarray}
%
Equations~(\ref{e2}) and (\ref{e3}) give a relation between the surface charge density $\sigma$,
the mid-plane potential $\psi_m$ and $d$, in terms of the Jacobi elliptic functions~\cite{abramowitz,Safinya},
\begin{eqnarray}
\label{e4}
\frac{\sigma}{e} = \frac{\kd}{4\pi\lb}\frac{m^2-1}{\sqrt{m}}\frac{{\rm sn}(u_s|m^2)}{{\rm cn}(u_s|m^2){\rm dn}(u_s|m^2)} \, ,
\end{eqnarray}
with $u_s \equiv d/(4\ld\sqrt{m})$ and $m \equiv \exp\left(\psi_m\right)$ as defined above.
For fixed surface charge, this relation gives the mid-plane potential, $\psi_m$.
Then, the disjoining pressure can be calculated from eq.~(\ref{e2a}).
When $d$ is the smallest length scale in the system, it can be shown
that the disjoining pressure in the so-called {\it ideal-gas} regime~\cite{Safinya,andelman2005,andelman1995} scales as,
\begin{eqnarray}
\label{e5}
\Pi_{_{\rm CC}} \simeq \frac{\kbt}{\pi\lb\lgc}\,\frac{1}{d} \sim d^{-1} \, .
\end{eqnarray}
The density (per unit volume) of the counter-ions is almost
constant between the two charged plates and is equal to $2|\sigma|/(ed)$.
This counter-ion density neutralizes the surface charge density, $\sigma$,
and the main contribution to the pressure comes from the entropy of an ideal-gas behavior of the counter-ion cloud.


In the second case of a CP boundary-condition $\psi_s$ is fixed, and unlike
the CC case, here the counter-ion concentration remains constant near
each of the planes, as it uniquely depends on the value of the surface potential, $\psi_s$, through the Boltzmann factor~\cite{banyaakov2010}.
The corresponding surface charge density, $\sigma$, is proportional to $d$ at small separation, $\sigma \sim d$.
Therefore, it vanishes for $d\to 0$.
Using the Taylor expansion of elliptic functions~\cite{abramowitz},
one can evaluate the leading terms in the surface potential for small $d$ as,
\begin{eqnarray}
\label{e6}
\psi_s \simeq \psi_m - \left(1 - m^2\right)u_s^2 \, .
\end{eqnarray}
Substituting the above equation into the disjoining pressure expression, eq.~(\ref{e2a}), yields to second order in $d$,
\begin{eqnarray}
\label{e7}
\nonumber \Pi_{_{\rm CP}} &\simeq& \kbt n_b \left(  4\sinh^2(\psi_s/2) -  \sinh^2(\psi_s)\frac{(\kd d)^2}{8} \right) \\
&\simeq& {const.} + {\cal O}\left(d^2\right) \, .
\label{CP1}
\end{eqnarray}
The above equation shows that the disjoining pressure for the CP boundary-condition
goes to a constant value, $\Pi_0$, for vanishing inter-plate separation, $d\to 0$,
with a leading correction term proportional to $d^2$.


\section{Single-site process}

As an example of a CR boundary-condition, we consider a surface that is composed of ionizable groups ({\it e.g.}, charged phospholipids).
Each group can release a counter-ion into the solution in a {\it single-site} dissociation process.
We first focus on such single-site CR dissociation process and refer to it as ${\rm CR_1}$. It is the simplest and most common CR process, and it will be extended below to multi-site processes.
The surface dissociation/association can be described by the reaction:
\begin{eqnarray}
\label{e8}
{\rm A}^+ + {\rm B}^- \rightleftarrows {\rm AB} \, ,
\end{eqnarray}
where A denotes a surface site that can be either ionized $({\rm A}^+)$
or neutral (AB).
The process of dissociation/association is characterized by an equilibrium constant $K_{\rm d}$
through the law of mass action
\begin{eqnarray}
\label{e9}
K_{\rm d} = \frac{[{\rm A}^+][{\rm B}^-]_{\rm s}}{[{\rm AB}]} \, ,
\end{eqnarray}
where $[{\rm A}^+]$, $[{\rm B}^-]_{\rm s}$ and [AB] denote the three corresponding surface concentrations.
The equilibrium condition of eq.~(\ref{e9}) can be written in terms of $\phi_s \equiv \sigma(\psi_s) a^2/e \sim [{\rm A}^+]$,
\begin{eqnarray}
\label{e10}
\nonumber\phi_s &=& \frac{1}{1 + \phi_b\e^{ - \alpha + \psi_s  }} \\
&=&   \half - \half\tanh\left[(\ln{\phi_b} - \alpha + \psi_s  )/2\right] \, ,
\end{eqnarray}
where $a^3$ is the ion volume, $\phi_b = a^3n_b$ is the ionic volume fraction
and we have introduced a surface interaction parameter $\alpha = \ln(a^3 K_{\rm d})$.
From $\sigma(\psi_s)$, eq.~(\ref{e4}) and eq.~(\ref{e6})
one obtains explicitly $\psi_m$.
By using the Taylor expansion of elliptic functions~\cite{abramowitz}
in eqs.~(\ref{e4}) and (\ref{e10}), it is clear that as $d \to 0$, $m$ diverges, but this divergency is weaker than $d^{-1}$.
It yields a diverging  ${\rm CR_1}$ disjoining pressure for small $d$,
\begin{eqnarray}
\label{e11}
\Pi^{(1)}_{_{\rm CR}} \simeq \sqrt{2}\kbt\e^{\alpha/2}a^{-5/2}d^{-1/2} \sim d^{-1/2}\, ,
\label{res1}
\end{eqnarray}
where the superscript in $\Pi^{(1)}_{_{\rm CR}}$ indicates that it corresponds to a  CR$_1$ process.
Note that just like $\Pi_{_{\rm CC}}$, $\Pi^{(1)}_{_{\rm CR}}$ does not depend on the bulk salt concentration, $n_b$.

\begin{figure}
\centering
\includegraphics[scale=0.7]{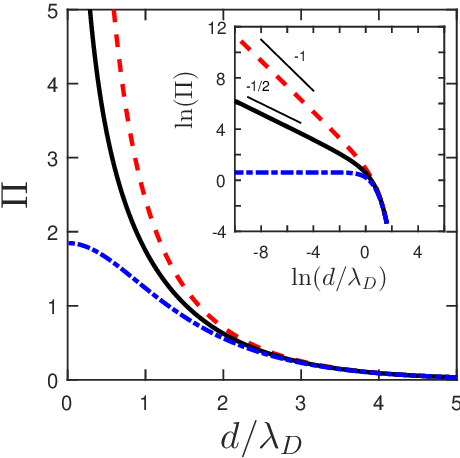}
\caption{ \textsf{ The dimensionless disjoining pressure, $\Pi$ [in units of $\kbt/(4\pi\lb\ld^2)$],
for the three boundary conditions: constant charge (CC - dashed red line), constant potential (CP - dotted-dashed blue line)
and charge regulation for {\it single-site} dissociation process (${\rm CR_1}$ - solid black line).
The pressure inequality, seen in the figure, $\Pi_{_{\rm CC}} \geq \Pi_{_{\rm CR}} \geq \Pi_{_{\rm CP}}$,
is an inequality that holds in general for the case of charge regulation consisting of a  single-site process.
The parameters used are: $a = 5 \, {\rm \AA}$, $n_b = 0.1\, {\rm M}$ and $\alpha = -6$ (${\rm pK} \simeq 1.48$).
In the inset we present the same disjoining pressure on a log-log plot, demonstrating its scaling with the inter-plate distance, $d/\ld$.
}}
\label{figure2}
\end{figure}

Another possible single-site process is the process of charging a neutral surface,
\begin{eqnarray}
\label{a1}
{\rm A} + {\rm B}^{+} \rightleftarrows {\rm AB^{+}} \, .
\end{eqnarray}
The equilibrium condition for this ${\rm CR_1}$ process can be written as,
\begin{eqnarray}
\label{a2}
\nonumber \phi_s &=& \frac{\phi_b}{\phi_b + \e^{\alpha + \psi_s}} \\
&=& \half - \half\tanh\left[(-\ln\phi_b + \alpha + \psi_s  )/2\right] \, .
\end{eqnarray}
Note the eq.~(\ref{a2}) reduces to eq.~(\ref{e10}) for the mapping:
$\phi_s\to 1-\phi_s$ and $\psi_s\to -\psi_s$.
Repeating the same procedure as above, we obtain a somewhat different disjoining pressure,
\begin{eqnarray}
\label{a3}
\Pi^{(1)}_{_{\rm CR}} \simeq \sqrt{2}\kbt n_b \e^{-\alpha/2}a^{1/2}d^{-1/2} \sim d^{-1/2} \, ,
\label{res2}
\end{eqnarray}
which also diverges as $d^{-1/2}$ but with a different prefactor that is linear
in the bulk concentration.
Note that these results do not depend on the sign of the surface site.

{
A typical pressure isotherm, $\Pi(d)$, is computed numerically for the ${\rm CR_1}$ process and is shown in Fig.~\ref{figure2}.
The mid-plane potential, $\psi_m$, is obtained as a function of inter-membrane separation, $d/\ld$,
using eqs.~(\ref{e2}), (\ref{e4}) and (\ref{e10}).
The pressure is calculated via eq.~(\ref{e2a}) and the surface potential from eq.~(\ref{e2}).
From the extrapolation of the surface-potential value at large
inter-plate separations,
we obtain the surface potential and the surface charge for CP and CC, respectively.}
The different pressure isotherms obey the inequality: $\Pi_{_{\rm CC}} \geq \Pi_{_{\rm CR}} \geq \Pi_{_{\rm CP}}$.
This is a general inequality that holds for charge regulations consisting of single-site dissociation process.
The log-log plot in Fig.~\ref{figure2} clearly shows the distinct $d^{-1/2}$ scaling for ${\rm CR_1}$,
confirming our analytical results, eqs.~(\ref{e11}) and (\ref{a3}).

An additional interesting observation can be made for the vanishing inter-plate separation, $d\to 0$.
The results for $\Pi^{(1)}_{_{\rm CR}}(d\to0)$ can be obtained from $\Pi_{_{\rm CC}}(d\to0)$ of eq.~(\ref{e5})
by substituting the surface charge, $\sigma(d)$, for the single-site process into the Gouy-Chapman length.
This shows a resemblance of the ${\rm CR}_1$ and CC processes, and gives some insight to the understanding of the different $\Pi^{(1)}_{_{\rm CR}}$ scalings.
The $\Pi^{(1)}_{_{\rm CR}}$ divergence is due to counter-ions that are bounded between the planes and neutralize the surface charge, as in the CC case.
However, the surface charge itself is not constant but decreases with $d$ as explained above.  Namely,
some of the counter-ions adsorb onto the surface in order to neutralize it.
Therefore, less counter-ions are bounded between the planes and the entropic penalty is reduced, as compare to the CC case.

The CR scaling results differ substantially from the disjoining pressure
of the CC $(\Pi_{_{\rm CC}} \sim 1/d)$ as well as CP $(\Pi_{_{\rm CP}} \sim {\rm const})$ boundary conditions.
There is a fundamental difference between the three boundary conditions,
making it clear that the disjoining-pressure scaling for small separations for the CC and CP boundary conditions
{\it cannot} be obtained by any limiting behavior of the CR boundary condition.

In previous works on charge regulation{~\cite{Borkovec1,Borkovec2,notalways1,Chan,Chan1}},
based on the same dissociation model,
additional approximations were used, including linearization of the CR boundary condition or linearization of the PB equation.
In these works, it has been shown that CR can reduce to CC or CP in different limits {of the differential capacitance}.
In contrast, our results show that the three disjoining pressures, $\Pi_{_{\rm CR}}$, $\Pi_{_{\rm CC}}$ and $\Pi_{_{\rm CP}}$ scale differently
in the small $d$ limit (as in eqs.~(\ref{res1}) and~(\ref{res2})), and point out that the CR case
does not generally reduce to the CC or CP ones.
It is not their generalization but rather a third distinct case of its own merit.


\section{Multi-site process}

{The inadequacy of the presumed limiting nature of CC and CP boundary conditions is equally apparent when one considers
more complicated surface dissociation/association processes that involve several ionic species.
In fact, although the disjoining pressure in this case is also bound between $\Pi_{\rm CC}$ and $\Pi_{\rm CP}$, it has a different scaling law than in the single-site CR process.} This
shows that CR has a rich behavior that depends on the number of surface dissociation/assciation processes.
As an illustrative example, we consider a dissociable surface with two independent dissociation/association processes
referred to as ${\rm CR_2}$ and described by:
\begin{eqnarray}
\label{e12}
\nonumber& &{\rm A}_1 + {\rm B}_1^+ \rightleftarrows {\rm A_1B_1^+} \, , \\
& &{\rm A}_2 + {\rm B}_2^- \rightleftarrows {\rm A_2B_2^-} \, ,
\end{eqnarray}
where ${\rm A_{1,2}}$ are two different surface binding sites.
A specific example for two dissociation processes can be~\cite{healy}:
\begin{eqnarray}
\label{e13a}
\nonumber{\rm A_1 H} + {\rm H}^+ &\rightleftarrows& {\rm A_1H_2^+} \, , \\
{\rm A_2H} + {\rm OH}^- &\rightleftarrows& {\rm A_2^- + H_2O} \, .
\end{eqnarray}
The equilibrium condition yields
\begin{eqnarray}
\label{e13}
\phi_s = \frac{p \phi_b}{\phi_b + \e^{\,\alpha_1 + \psi_s}} - \frac{(1-p)\phi_b}{\phi_b + \e^{\,\alpha_2 - \psi_s}} \, ,
\end{eqnarray}
where $\alpha_{1,2}$ are two surface interaction parameters for dissociation/association of ${\rm B}_{1,2}$,
and $p = N_{1}/N$ is the surface fraction of A$_1$ sites with $N$ being the total number of sites and $N_{1}$ the number of the ${\rm A}_{1}$ sites.
For $p=1$ (or $p=0$), eq.~(\ref{e13}) reduces to eq.~(\ref{a2}) (or a similar equation for negative binding ions)
and the single-site process case (${\rm CR_1}$) is recovered.
Note that bulk electro-neutrality dictates the equality of the bulk concentration of ${\rm B}_{1}^+$ and ${\rm B}_{2}^-$.

Without loss of generality we can focus on the situation in which $\alpha_2 \to -\infty$, {\it i.e.}, strong
adsorption of the ${\rm B}_{2}^-$ ions, giving the approximate form of eq.~(\ref{e13}) (similar to eq.~(6) in Ref. \cite{Borkovec1}),
\begin{eqnarray}
\label{e14}
\phi_s \simeq \frac{p \phi_b}{\phi_b + \e^{\,\alpha_1 + \psi_s}} - (1-p)  \, ,
\end{eqnarray}
such that the adsorption of ${\rm B}_{2}^-$ ion is similar to a constant surface charge that remains fixed.

We repeat the same steps as done above for the ${\rm CR_1}$ process in order
to derive the limiting form of the disjoining pressure for $d \to 0$, and obtain, to first order in $d$,

\begin{eqnarray}
\label{e15}
\nonumber \Pi^{(2)}_{_{\rm CR}} \!\!\!\! &\simeq& \!\!\!\!\kbt n_b  \frac{(m_0 - 1)^2 }{m_0} \Bigg[ 1 - d \cdot \frac{n_b a^2}{2}
\frac{ p(m_0 + 1)^2}{(1-p)(2p-1)m_0} \Bigg] \\
&\simeq& \Pi_0 - \Pi_1 d \, ,
\end{eqnarray}
where $m = m_0 + m_1d + \ldots$, and
\begin{eqnarray}
\label{e15a}
m_0 = \frac{2p - 1 }{1 - p}n_ba^3\e^{-\alpha}  \, ,
\end{eqnarray}
is the first term in the expansion of $m=\exp(\psi_m)$.
This result is similar to the CP result as the disjoining pressure goes to a constant value for $d \to 0$,
but the first correction in ${\rm CR_2}$ is linear in $d$, unlike the first CP correction that scales as $d^2$.
This pressure expression is valid for $0.5 < p < 1$, while for $p \to 1/2$,
$\Pi^{(2)}_{_{\rm CR}} \sim d^{-1/2}$ as for the ${\rm CR_1}$ case.
For smaller $0< p < 0.5$, there are always some fixed surface charges and the pressure expression for $d\to0$
reduces to the one of CC, eq.~(\ref{e5}), with $|\sigma|/e = 1 - 2p$.
Note that these expressions hold in the limit $\alpha_2 \to -\infty$,
while for any finite $\alpha_2$, the pressure expression always has the same limiting behavior as eq.~(\ref{e15}),
$\Pi^{(2)}_{_{\rm CR}} \simeq C_0 - C_1 d$, but with different coefficients, $C_0$ and $C_1$.

\begin{figure}
\centering
\includegraphics[scale=0.7]{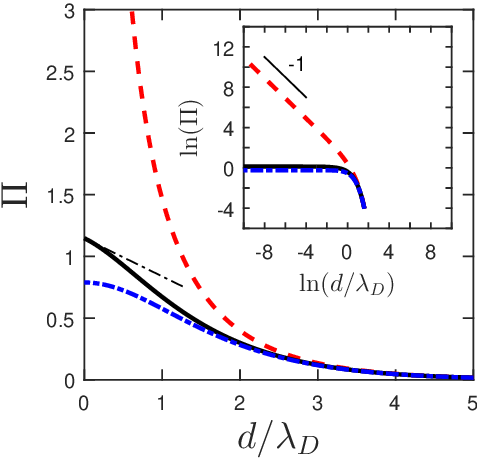}
\caption{ \textsf{ The dimensionless disjoining pressure, $\Pi$ [in units of $\kbt/(4\pi\lb\ld^2)$],
for the three different boundary conditions: constant charge (CC - dashed red line), constant potential (CP - dotted-dashed blue line)
and charge regulation for {\it two-site} dissociation process (${\rm CR_2}$ - solid black line).
The dotted-dashed thin (black) line indicates the non-zero slope of the ${\rm CR_2}$ at $d=0$,
and the parameters used are as in Fig.~\ref{figure2} with $p=0.7$ ($m_0 > 1$).
In the inset, we present the same disjoining pressures on a log-log plot, demonstrating their
scaling with the inter-plate distance, $d/\ld$.
}}
\label{figure3}
\end{figure}

{A typical pressure isotherm, $\Pi(d)$, is computed numerically for the ${\rm CR_2}$ process and is shown in Fig.~\ref{figure3}.
The calculation is exactly the same as for Fig.~\ref{figure2}, but instead of eq.~(\ref{e10}), we use eq.~(\ref{e14}),
which corresponds to the CR$_2$ boundary condition.
The log-log plot (inset) clearly  shows that $\Pi(d)$ tends towards a constant value, $\Pi_0$, as $d\to 0$,
with a constant negative slope, $\Pi_1$ (dotted-dashed thin black line), as derived in eq.~(\ref{e15}).}

The same calculation can be performed for any {\it multi-site} dissociation processes ${\rm CR}_{n\geq 2}$.
It can be shown that
$\Pi^{(n\geq2)}_{_{\rm CR}}(d\to 0) \simeq C_0^{(n)} - C_1^{(n)} d$,
where $C_{0,1}^{(n)}$ are the two coefficients in the small $d$ expansion, whose value depends on $n\geq 2$.
The value of $m_0$ also depends on $n\geq 2$ and is determined by examining the vanishing $\phi_s$ limit
in the equilibrium condition [eq.~(\ref{e13}) for the ${\rm CR_2}$ dissociation process].
This pressure scaling is a result of the competition between the two (or more) dissociation/association processes of anions and cations.
Unlike the  CC and ${\rm CR_1}$ cases, where counter-ions have to stay bounded between the planes to neutralize the surface charge,
in the multi-site process the planes are neutralized by the two (ot more) competing processes.
Therefore, no counter-ions remain between the plane and there is no entropic penalty.

\section{Conclusions}

In this Letter, we have shown that the CR boundary condition implies a much richer behavior than just an interpolation between
the limiting forms of the CC and CP boundary conditions.
Our conclusions are based on the full non-linear PB equation, as well as the non-linear form of
the charge regulation conditions.
They differ from previous claims that are based on linearization schemes{~\cite{Borkovec1,Borkovec2,notalways1,Chan,Chan1}}.

{We have shown that for both single-site (${\rm CR_1}$) and multi-site (${\rm CR_{n\geq2}}$) surface dissociation/association processes
the disjoining pressure is indeed bounded by the CC and CP limits,
while its scaling for small separations depends on the process type and, generally, is at odds with both the CC and CP limiting cases.}
This is clear from the different scaling of the single-site process (${\rm CR_1}$),
$\Pi^{(1)}_{_{\rm CR}}$ that scales as $d^{-1/2}$, while $\Pi_{_{\rm CC}}$ scales as $d^{-1}$ and $\Pi_{_{\rm CP}}$ tends to a constant.
We note that all considered boundary conditions lead to an identical (universal) separation scaling for large $d$.

The single-site case is more similar to the CC boundary condition as it diverges for small separations.
As was explained above, the pressure isotherm, $\Pi^{(1)}_{\rm CR}$,
can be obtained by substituting the surface charge density, $\sigma(d)$,
into the Gouy-Chapman length of the disjoining pressure expression, $\Pi_{_{\rm CC}}$ of eq.~(\ref{e5}),
for small $d$.
Furthermore, for multi-site dissociation processes, we have shown that $\Pi^{(n)}_{_{\rm CR}}$ ($n\geq 2$)
is similar to the CP case, as it tends towards a constant value for small separations, $\Pi^{(n)}_{_{\rm CR}} \simeq C_0^{(n)} - C_1^{(n)} d$.
Nevertheless, as is apparent from its negative slope at $d=0$, it differs from the CP case whose slope is zero.

In summary, the CR process is shown to be a distinct type of boundary condition with particular scaling
behavior,
and cannot be considered as a generalization of the CC and CP cases.
One should also keep in mind the fundamental difference between the CR single-site process and the multi-site ones.

\acknowledgments
{\bf Acknowledgements.~~~}
We thank F. Mugele for discussions that push-started this research
and to M. Biesheuvel, N. Boon, M. Borkovec, A. Cohen, Y. Nakayama, R. Netz, S. Safran, and especially R. van Roij
for useful comments and numerous suggestions.
This work was supported in part by the Israel
Science Foundation (ISF) under Grant No. 438/12,
the US-Israel Binational Science Foundation (BSF) under
Grant No. 2012/060.
R.P. would like to thank the Slovene research agency ARRS for support through Grant No. P1-0055.



\end{document}